\documentclass{PoS}

\usepackage{amsmath,amsfonts,amssymb}
\usepackage{breqn}
\usepackage{cite}
\usepackage{comment}
\usepackage[usenames,dvipsnames]{xcolor}
\definecolor{darkgreen}{RGB}{34,139,34}

\title{Recent new methods and applications of the differential equation approach to master integrals}

\ShortTitle{Recent new applications of differential equations to master integrals}

\author{Costas G. Papadopoulos, Damiano Tommasini and \speaker{Christopher Wever}
        \thanks{Supported by the Research Funding Program ARISTEIA, HOCTools (co-financed by the European 
         Union (European Social Fund ESF) and Greek national funds through the Operational Program 
         "Education and Lifelong Learning" of the National Strategic Reference Framework (NSRF))}\\
        Institute of Nuclear Physics, NCSR "Demokritos", Athens, Greece\\
        E-mail: \email{costas.papadopoulos@cern.ch},
                \email{tommasini@inp.demokritos.gr},
                \email{wever@inp.demokritos.gr}}

\abstract{A short review is given of the simplified differential equations approach to Master Integrals, which was 
recently proposed by one of the authors. We show its applicability by calculating some non-trivial two-loop planar
Master Integrals, namely those contributing to amplitudes of massive diboson $VV'$ production at the LHC with 
massless internal lines.}

\FullConference{ Loops and Legs in Quantum Field Theory - LL 2014,\\
                 27 April - 2 May 2014 \\
                 Weimar, Germany }

\begin{document}

\section{Introduction}

At the first run of the LHC, proton collisions have reached the record-setting high energies of 8 TeV. The second run 
of the LHC, which is expected to start in 2015, will push the energy and luminosity even higher. In order to keep up with the increasing experimental accuracy as more data is collected, 
more precise theoretical predictions and higher loop calculations will be required.

With the better understanding of reduction of one-loop amplitudes to a set of Master Integrals ({\bf MI}) based on unitarity methods~\cite{Bern:1994cg,Bern:1994zx} 
together with their implementation at the integrand level via the OPP method~\cite{Ossola:2006us,Ossola:2008xq}, one-loop calculations have been 
fully automated in many numerical tools~\cite{vanHameren:2009dr,Berger:2009zg,Bern:2013gka,Bredenstein:2009aj,Bevilacqua:2009zn,Bevilacqua:2010ve}. In recent years, a lot of progress has been made towards the extension of on-shell 
reduction methods to the two-loop order at the integral~\cite{Gluza:2010ws,Kosower:2011ty} as well as the integrand~\cite{Mastrolia:2011pr,Badger:2012dp,Badger:2013gxa} level. Contrary to the MI at 
one-loop, which have been known for a long time already~\cite{'tHooft:1978xw}, a complete library of MI at two-loops is still missing. 

Starting from the works of~\cite{Goncharov:1998aa,Remiddi:1999ew,Goncharov:2001aa}, there has been a building consensus that the so-called {\it Goncharov Polylogarithms} ({\bf GPs}) form a 
functional basis for many MI. A very fruitful method for calculating MI and expressing them in 
terms of GPs is the differential equations ({\bf DE}) approach~\cite{Kotikov:1990kg,Remiddi:1997ny}, which has been used in the 
past two decades to calculate various MI at two-loops~\cite{Caffo:1998du,Gehrmann:1999as,Gehrmann:2000zt,Gehrmann:2001ck,Martin:2003qz,Birthwright:2004kk,Awramik:2004ge,Aglietti:2004ki,Anastasiou:2006hc,Bonciani:2008wf,Henn:2014lfa,Caola:2014lpa}. In~\cite{Papadopoulos:2014lla} a variant of the traditional DE approach 
to MI was presented, which was coined the Simplified Differential Equations ({\bf SDE}) approach. In this 
paper a short review of the SDE approach and an application to the four-point 
two-loop planar MI with two different massive external legs and massless internal propagators is presented~\cite{Papadopoulos:2014b}.

\section{Simplified differential equations approach}

\begin{figure}[t!]
\centering
\includegraphics[width=0.4 \linewidth]{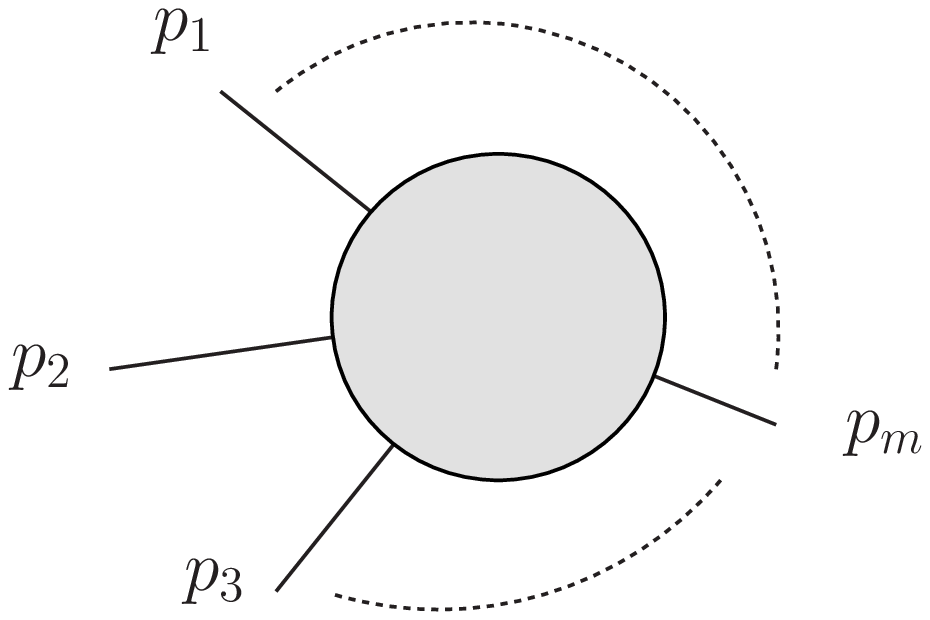} \hspace{1.5 cm}
\includegraphics[width=0.4 \linewidth]{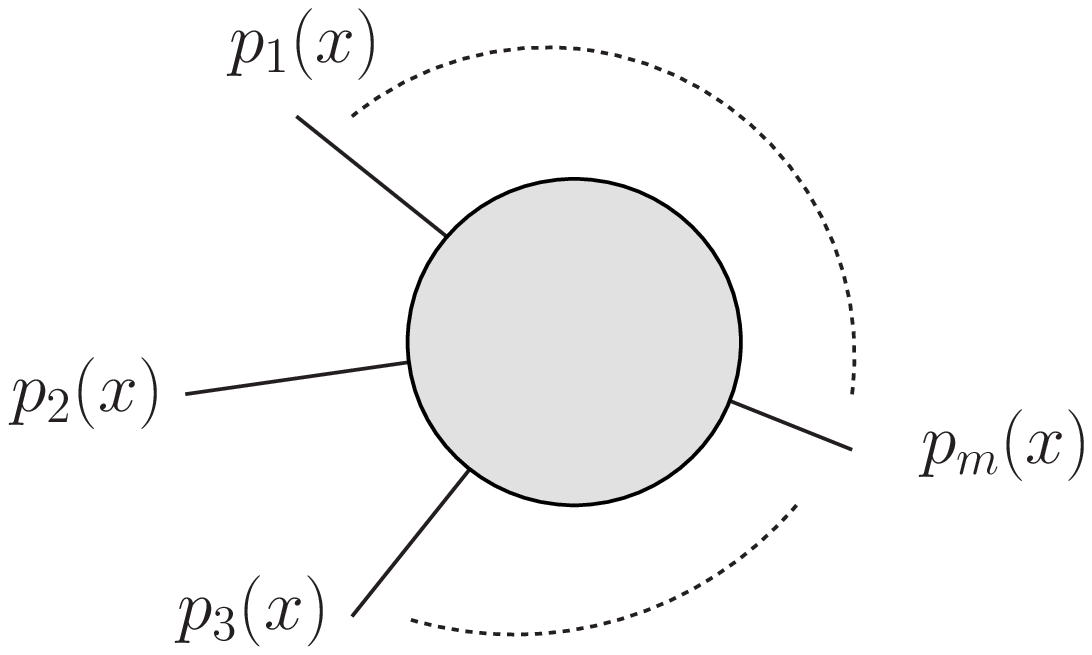}
  \caption{General graphs represented by (\protect\ref{eq:loopgen}) on the left and (\protect\ref{eq:loopgenx}) on the right.}
  \label{fig:xparam}
\end{figure}

The SDE approach was introduced in~\cite{Papadopoulos:2014lla} and its main points are reviewed in this section. Assume one is 
interested in calculating an $l-$loop Feynman integral whose graph with external incoming momenta $\{p_j\}$ is shown on the left 
hand side in Figure \ref{fig:xparam}. The method has been developed for massless internal lines and therefore all propagators are 
massless in the rest of this paper. All relevant Feynman integrals are a subset of the following class of 
loop integrals:
\begin{equation}
G_{a_1\cdots a_n}(\{p_j\},\epsilon)=\int\left(\prod_{r=1}^l d^dk_r\right)\frac{1}{D_1^{2a_1}(k,p)\cdots D_n^{2a_n}(k,p)}, \hspace{0.5 cm}
D_i(k,p)=c_{ij}k_j+d_{ij}p_j,
\label{eq:loopgen}
\end{equation}
where the denominators are defined in such a way that all scalar product invariants can be written as a linear combination of them. 
The exponents $a_i$ are integers and may be negative in order to generate numerators.

The above class of Feynman integrals satisfy so-called {\it integration by parts} ({\bf IBP}) identities~\cite{Chetyrkin:1981qh,Tkachov:1981wb} in dimensional 
regularization ({\bf DR}) with $d=4-2\epsilon$:
\begin{equation}
\int\left(\prod_{r=1}^l d^d k_r\right) \frac{\partial}{\partial k_j^{\mu}}
\left(\frac{v^{\mu}}{D_1^{2a_1}D_2^{2a_2}\cdots D_n^{2a_n}}\right)\stackrel{DR}{=}0, \label{eq:IBP}
\end{equation}
where the vector $v_{\mu}$ is in practice taken to be one of the loop or external momenta. By solving the above identities 
one can reduce any integral $G_{a_1\cdots a_n}$ as a linear combination of MI with rational coefficients in the scalar products 
and space-time dimension $d$. Reduction by IBP is by now implemented in public tools~\cite{Smirnov:2008iw,vonManteuffel:2012np}.

If a basis vector of MI $\vec{G}^{MI}(s)=\{G_{b_1\cdots b_n}|(b_1\cdots b_n)\in\text{Master Integrals}\}$ is known, then any integral $G_{a_1\cdots a_n}$ 
may be calculated after IBP reduction. In the traditional DE method $\vec{G}^{MI}(s)$ is differentiated with respect to a Lorentz invariant (or set of 
invariants) $s$ and the resulting integrals are reduced by IBP to give a linear system of DE for 
$\vec{G}^{MI}(s)$~\cite{Kotikov:1990kg,Remiddi:1997ny}. The invariant $s=f(\{p_i.p_j\})$ that is differentiated to is in 
general a function of the scalar products and is defined on a case by case basis. 

It has been conjectured~\cite{Henn:2013pwa} that it is always possible to make a rotation of the vector $\vec{G}^{MI}(s)$ to get a DE 
of the following {\it canonical} form:
\begin{equation}
\text{DE:} \hspace{1 cm} \frac{\partial}{\partial s} \vec{G}^{MI}(s,\epsilon)
\stackrel{IBP}{=}\epsilon\overline{\overline{M}}(s)\cdot\vec{G}^{MI}(s,\epsilon), \hspace{0.5 cm} s=f(\{p_i.p_j\}). \label{eq:DE}
\end{equation}
In this new basis the MI may be directly solved as exponentiated integrals of the matrix $\overline{\overline{M}}$. For a large 
class of MI these integrals are expressible in terms of GPs, which are an iterative class of functions that generalize the usual 
logarithms and polylogarithms~\cite{Goncharov:1998aa,Goncharov:2001aa}:
\begin{gather}
GP(\underbrace{\alpha_1,\cdots ,\alpha_n}_{\text{weight n}};x):=\int_0^x dx' \frac{GP(\alpha_2,\cdots ,\alpha_n;x')}{x'-\alpha_1}, \nonumber\\
GP(;x)=1, \ \ \ \ GP(\underbrace{0,\cdots ,0}_{\text{n times}};x)=\frac{1}{n!}\log^n(x), \label{eq:GPs}
\end{gather}
where in general $\alpha_i,x\in \mathbb{C}$. The subsequent expansion of $\vec{G}^{MI}$ in $\epsilon$ results in a power series whose coefficients are 
uniform in the {\it weight} of the GPs~\cite{Henn:2013pwa}.

In the SDE approach~\cite{Papadopoulos:2014lla} the MI are differentiated with respect to an {\it externally introduced parameter} that will 
be denoted by $x$. As shown on the right hand side in Figure \ref{fig:xparam}, the external incoming momenta are now {\it parametrized} 
linearly in terms of $x$ as $p_i(x)=p_i+(1-x)q_i$, where the $q_i$'s are a linear combination of the original external momenta $\{p_i\}$ such 
that $\sum_iq_i=0$. If $p_i^2=0$, the parameter $x$ captures the off-shellness of the massive external legs. 
The class of Feynman integrals in (\ref{eq:loopgen}) are now dependent on $x$ through the external momenta:
\begin{equation}
G_{a_1\cdots a_n}(x,s,\epsilon)=\int\left(\prod_{i=1}^l d^dk_i\right)\frac{1}{D_1^{2a_1}(k,p(x))\cdots D_n^{2a_n}(k,p(x))}, 
\hspace{0.5 cm} s=\{p_i.p_j\}|_{i,j}, \label{eq:loopgenx}
\end{equation}
where contrary to the traditional DE approach, the Lorentz invariants $s$ are here defined as the usual scalar products. 
Note that as $x\rightarrow 1$, the original configuration of the loop integrals (\ref{eq:loopgen}) are reproduced. The vector 
$\vec{G}^{MI}(x)$ is now dependent on $x$ and one differentiates it with respect to $x$ to get a linear system of 
differential equations:
\begin{equation}
\text{SDE:} \hspace{1 cm} \frac{\partial}{\partial x} \vec{G}^{MI}(x,s,\epsilon)
\stackrel{IBP}{=}\overline{\overline{M}}(x,s,\epsilon)\cdot\vec{G}^{MI}(x,s,\epsilon), \ \ \ \ s=\{p_i.p_j\}|_{i,j}. \label{eq:DEx}
\end{equation}

\begin{figure}[t!]
\centering
\includegraphics[width=0.32 \linewidth]{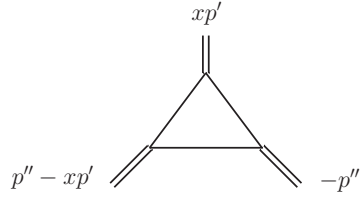}
  \caption{Required parametrization for massive triangles after possible pinchings of internal line(s).}
  \label{fig:xparam-tr}
\end{figure}

The MI with least amount of denominators $m_0$ are two-point integrals which can be easily calculated analytically with other 
methods. Furthermore, because of the form of the IBP identities (\ref{eq:IBP}) the DE of MI with $m$ denominators only depend on MI 
with at most $m$ denominators. This structure of the DE makes it possible to first solve the MI with $m_0+1$ denominators, then those 
with $m_0+2$ denominators and so forth. In other words, in practice the DE may be solved from the {\it bottom-up}. 
For many cases, the MI with $m_0$ denominators are expressible in terms of GPs (\ref{eq:GPs}) and one needs to choose the 
parametrization of the external momenta in $x$ such that this GP-structure for the MI with $m>m_0$ 
denominators holds as well. For the cases that we considered it was enough for us to choose the parametrization of the external 
legs such that after pinching internal lines the resulting triangles with three off-shell legs, if they appear, have the form 
given in Figure \ref{fig:xparam-tr}. In other words one of the external momentum should scale linearly with $x$, and another one 
should be independent of $x$.

As was noted in~\cite{Papadopoulos:2014lla}, in the SDE approach the {\it boundary terms} when solving from the bottom-up are almost always naturally captured by the singularities in the SDE themselves at $x=0$, which is precisely the lower integration boundary of the GPs. In this way, the SDE method is well suited for directly and efficiently expressing the MI in terms of GPs. In particular, the boundary terms of all the two-loop planar graphs given in the next section as well as those for many other examples we have considered are captured by the SDE without the addition of any constants in $x$.

\section{Application: Master integrals of the two-loop four-point planar amplitudes with two different massive external legs}

In this section a non-trivial application of the SDE approach and partial results of~\cite{Papadopoulos:2014b} are presented. We 
are interested in calculating the MI of two-loop QCD amplitude corrections contributing to massive pair production at the LHC, where the 
two outgoing particles may have different masses:
\begin{equation}
pp'\rightarrow V_1V_2, \ \ m_{V_1}\neq m_{V_2}\neq 0. \label{eq:ppVV}
\end{equation}
Both the planar and non-planar diagrams have already been calculated with the traditional DE method~\cite{Henn:2014lfa,Caola:2014lpa}. As a comparison of 
similar applicability, the {\it planar} diagrams were calculated with the SDE method in~\cite{Papadopoulos:2014b}. These results are necessary in order to calculate 
two-loop QCD corrections to $WZ$ production at the LHC or in general for off-shell particle pair production.

\begin{figure}[t!]
\centering
\includegraphics[width=0.29 \linewidth]{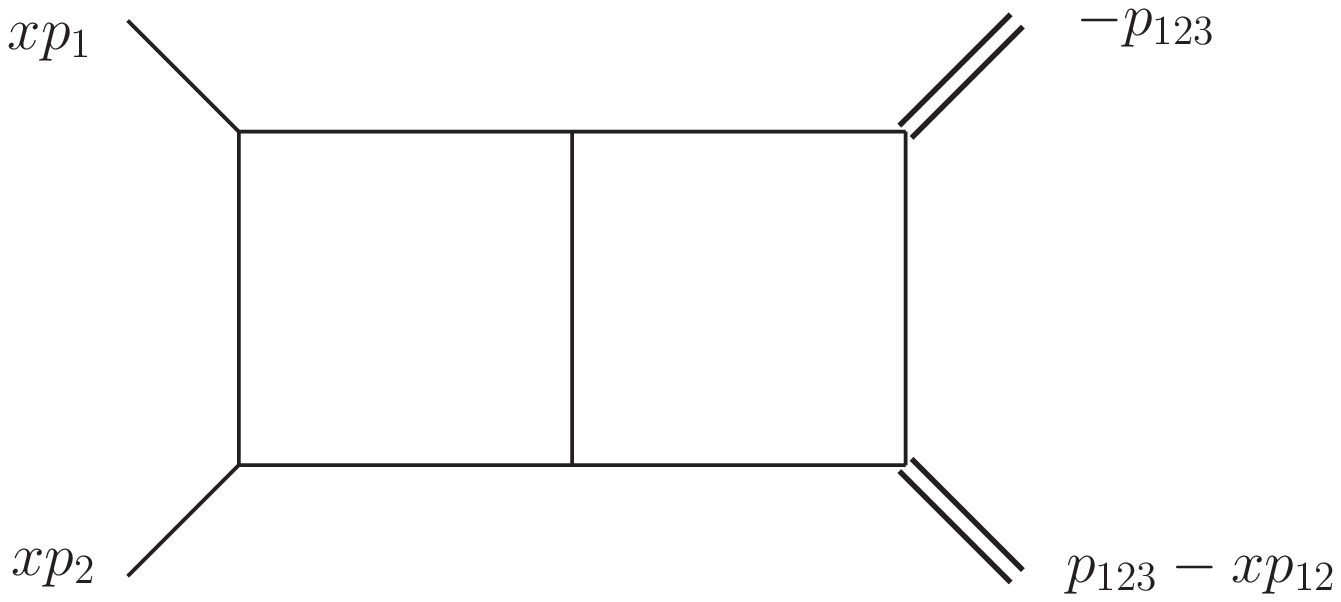} \hspace{0.5 cm}
\includegraphics[width=0.31 \linewidth]{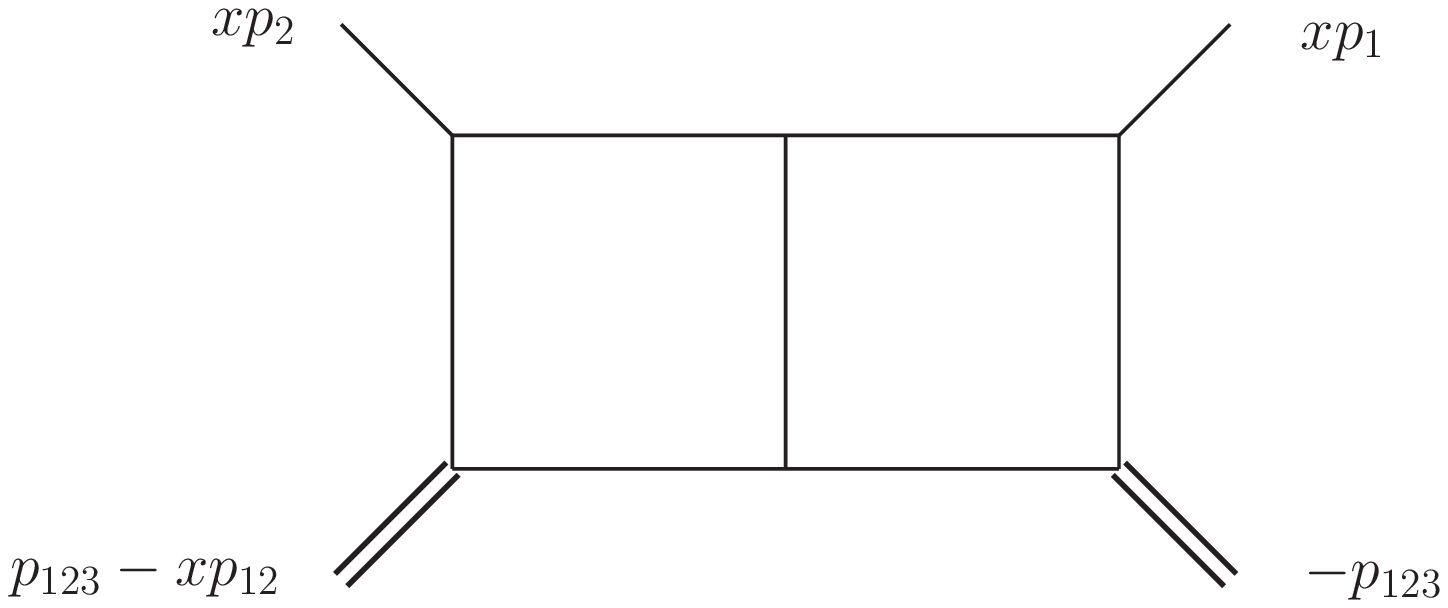} \hspace{0.5 cm}
\includegraphics[width=0.31 \linewidth]{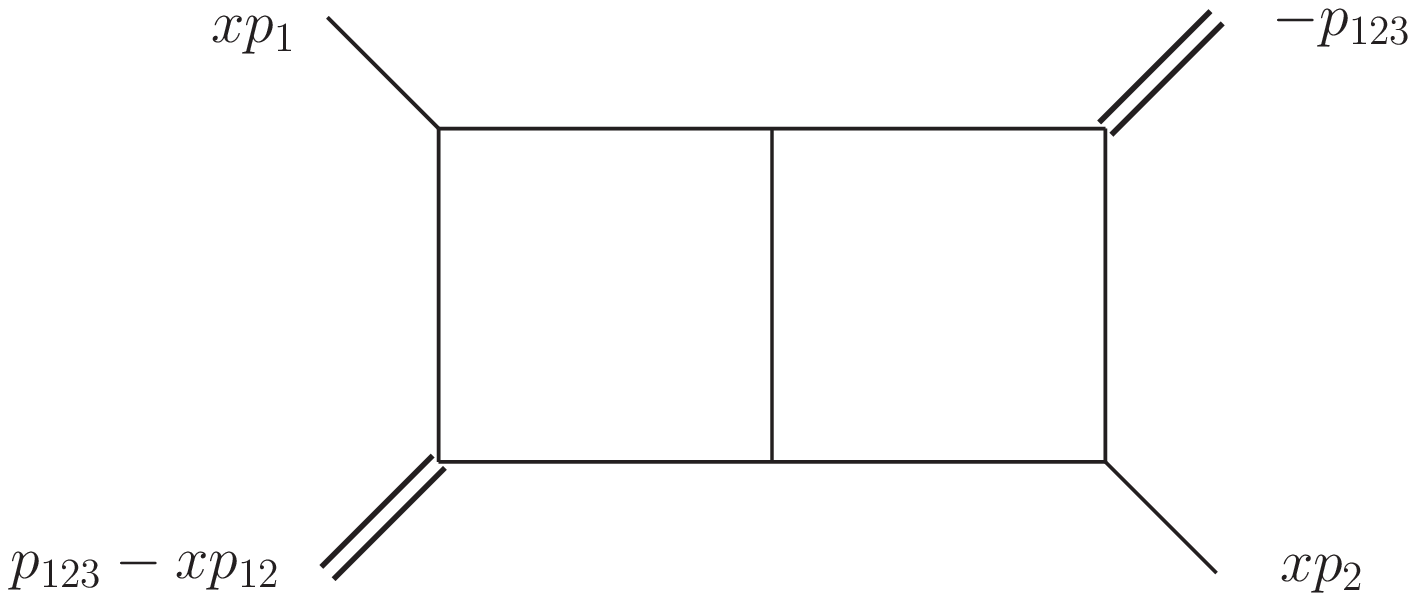}
  \caption{The parametrization of external momenta for the three planar double boxes of the families $P_{12}$ (left), $P_{13}$ (middle) and $P_{23}$ (right) 
    contributing to pair production at the LHC.}
  \label{fig:xparam-P}
\end{figure}

There are three families of planar integrals whose MI with maximum amount of denominators are graphically shown in Figure \ref{fig:xparam-P}. 
The three families are denoted by $P_{12},P_{13}$ and $P_{23}$ as was done in~\cite{Henn:2014lfa} and have 31, 29 and 28 MI respectively. In order to parametrize the external momenta 
we kept in mind Figure \ref{fig:xparam-tr}. In particular, the parametrization of $P_{12}$ and $P_{13}$ were chosen 
such as to satisfy the requirement of having massive triangles of the form shown in Figure \ref{fig:xparam-tr} after pinching 
the internal line(s) between the two massless legs. The parametrization of $P_{23}$ was then found by permuting the external 
legs accordingly. The class of loop integrals describing the families $P_{12}$ and $P_{13}$ are explicitly expressed in $x$ as:
\begin{eqnarray}
G^{P_{12}}_{a_1\cdots a_9}(x,s,\epsilon)&:=&\int \frac{d^dk_1}{i\pi^{d/2}}\frac{d^dk_2}{i\pi^{d/2}}
\frac{1}{k_1^{2a_1} (k_1 + x p_1)^{2a_2} (k_1 + x p_{12})^{2a_3} (k_1 + p_{123})^{2a_4}} \nonumber\\
&\times& \frac{1}{k_2^{2a_5} (k_2 - x p_1)^{2a_6} (k_2 - x p_{12})^{2a_7} 
(k_2 - p_{123})^{2a_8} (k_1 + k_2)^{2a_9}}, \label{eq:P12x}\\
G^{P_{13}}_{a_1\cdots a_9}(x,s,\epsilon)&:=&\int \frac{d^dk_1}{i\pi^{d/2}}\frac{d^dk_2}{i\pi^{d/2}}
\frac{1}{k_1^{2a_1} (k_1 + x p_1)^{2a_2} (k_1 + x p_{12})^{2a_3} (k_1 + p_{123})^{2a_4}} \nonumber\\
&\times& \frac{1}{k_2^{2a_5} (k_2 - x p_1)^{2a_6} (k_2 - p_{12})^{2a_7} 
(k_2 - p_{123})^{2a_8} (k_1 + k_2)^{2a_9}}. \label{eq:P13x}
\end{eqnarray}
Similarly, the class of integrals for the family $P_{23}$ equals:
\begin{eqnarray}
G^{P_{23}}_{a_1\cdots a_9}(x,s,\epsilon)&:=&\int \frac{d^dk_1}{i\pi^{d/2}}\frac{d^dk_2}{i\pi^{d/2}}
\frac{1}{k_1^{2a_1} (k_1 + x p_1)^{2a_2} (k_1 + p_{123}-x p_2)^{2a_3}
 (k_1 + p_{123})^{2a_4}} \nonumber\\
&\times& \frac{1}{k_2^{2a_5} (k_2 - p_1)^{2a_6} (k_2 + x p_2 - p_{123})^{2a_7} 
(k_2 - p_{123})^{2a_8} (k_1 + k_2)^{2a_9}}. \label{eq:P23x}
\end{eqnarray}

With the above parametrization, the solutions of the DE are all expressed in terms of GPs (\ref{eq:GPs}). The DE may be solved via the 
bottom-up approach and in this way one finds that all boundary terms are included by integrating the singularities at the boundary $x=0$ in the DE. For each family $P_i$, the first set of 
arguments $AG(P_i)$ appearing in the GPs, the $\alpha_j$'s in (\ref{eq:GPs}), are as follows:
\begin{gather}
AG(P_{12})=\left\{0,1,\frac{m_4}{s_{12}}, \frac{m_4}{m_4 - s_{23}}, \frac{s_{23}}{s_{12}} + 1\right\}, \nonumber\\
AG(P_{13})=AG(P_{12})\cup \left\{\xi_+,\xi_- ,
\frac{m_4 (m_4 - s_{23})}{m_4^2-s_{23} \left(m_4+s_{12}\right)}\right\} \!, \xi_{\pm}=\frac{m_4 s_{12} \pm \sqrt{m_4 s_{12} s_{23} (-m_4 + s_{12} + s_{23})}}{m_4 s_{12} - s_{12} s_{23}}, \nonumber\\
AG(P_{23})=AG(P_{12})\cup \left\{\frac{m_4-s_{23}}{s_{12}},\frac{m_4}{s_{12} + s_{23}}\right\}. \label{eq:alphabet}
\end{gather}
Note that the parameter $x$ does not appear in the first set of arguments of the GPs, but appears instead as the last argument as we integrate the DE. This can be for example explicitly seen 
in the solution of the scalar double box of the family $P_{13}$:
\begin{gather*}
G^{P_{13}}_{011111011}(x,s,\epsilon)=\frac{A_3(\epsilon)}{x^2 s_{12} (-m_4+x(m_4-s_{23}))^2}\left\{\frac{-1}{2\epsilon^4}
+\frac{1}{\epsilon^3}\!\left(\!-GP\left(\frac{m_4}{s_{12}};x\right)\!+2 \ \! GP\!\left(\!\frac{m_4}{m_4-s_{23}};x\!\right) \right.\right. \nonumber\\
\left.+2 \ GP(0;x)-GP(1;x)+\log \left(-s_{12}\right)+\frac{9}{4}\right)+\frac{1}{4\epsilon^2}\left(18 \ GP\left(\frac{m_4}{s_{12}};x\right)
-36 \ GP\left(\frac{m_4}{m_4-s_{23}};x\right) \right. \nonumber\\
\left.-8 \ GP\left(0,\frac{m_4}{s_{12}};x\right)+16 \ GP\left(0,\frac{m_4}{m_4-s_{23}};x\right)+8 \ GP\left(\frac{s_{23}}{s_{12}}
+1,\frac{m_4}{m_4-s_{23}};x\right)+\cdots\right) \nonumber\\
+\frac{1}{\epsilon}\left(9\left(GP\left(0,\frac{m_4}{s_{12}};x\right)+GP(0,1;x)\right)-4\left(GP\left(0,0,\frac{m_4}{s_{12}};x\right)+GP(0,0,1;x)\right)+\cdots\right) \nonumber\\
\left. +6 \left(GP\left(0,0,1,\xi_-;x\right)+GP\left(0,0,1,\xi_+;x\right)\right)
\!-\!2 \ \! GP \! \left(0,0,\frac{m_4}{m_4-s_{23}},\frac{m_4 \left(m_4-s_{23}\right)}{m_4^2-s_{23} \left(m_4+s_{12}\right)};x\right)
\! +\cdots\right\}.
\end{gather*}
The above MI is expressed in terms of a common factor $A_3(\epsilon)$, the parameter $x$ and the scalar products that are related 
to the Mandelstam variables $S,T$ and particle masses $M_3,M_4$ as follows:
\begin{gather}
S=s_{12}x^2, \hspace{0.5 cm} T=m_4 - (s_{12} + s_{23}) x, \hspace{0.5 cm} M_3^2=(1 - x) (m_4 - s_{12} x), \hspace{0.5 cm} M_4^2=m_4, \nonumber\\
s_{12}=p_{12}^2, \hspace{0.5 cm} s_{23}=p_{23}^2, \hspace{0.5 cm} m_4=p_{123}^2, \hspace{0.5 cm} A_3(\epsilon)=-\frac{\Gamma(1-\epsilon)^3\Gamma(1+2\epsilon)}
{\Gamma(3-3\epsilon)}. \label{eq:Pinv}
\end{gather}

As one may notice from the example $G^{P_{13}}_{011111011}$ above, the solutions are in general not uniform in the weight of the GPs~\cite{Henn:2013pwa}. The Goncharov polylogs can be numerically evaluated 
with the Ginac\footnote{We would like to thank S. Weinzierl for his help with Ginac.} library~\cite{Vollinga:2004sn}. We have not tried to simplify our analytical expressions by the use of 
symbol and coproduct techniques~\cite{Duhr:2011zq,Duhr:2012fh}, as we were mostly interested in showing the applicability of the SDE method, but 
this is expected to be possible if required. All the MI solutions of the three relevant families have been numerically compared in the Euclidean region for 
various phase points with the numerical program Secdec\footnote{We are thankful for the help of G. Heinrich and S. Borowka with Secdec.}~\cite{Carter:2010hi,Borowka:2012yc} and good agreement was found. For example, the numerical output of the 
analytical solution of the scalar double box $G^{P_{13}}_{011111011}$ at a Euclidean phase space point $x=1/3,s_{12}=-2,s_{23}=-5,m_4=-9$ equals:
\begin{gather*}
G^{P_{13}}_{011111011}(S=-2/9,T=-20/3,M_3^2=-50/9,M_4^2=-9)=-\frac{0.0191399}{\epsilon^4}-\frac{0.0292887}{\epsilon^3} \nonumber\\
+\frac{0.0239971}{\epsilon^2}+\frac{0.340233}{\epsilon}+0.870356+\mathcal{O}(\epsilon).
\end{gather*}
The above numerical expression has been compared with Secdec to 6 digits. For phase space points in the physical region great care 
needs to be taken of the Feynman $i\epsilon_F$ prescription related to the propagators. We refer to~\cite{Papadopoulos:2014b} 
for the complete discussion of how our solutions may be analytically continuated to the physical region. Furthermore, the numerical 
comparisons at Euclidean as well as physical phase space points for all double boxes are given in~\cite{Papadopoulos:2014b}, which 
will also contain all analytical solutions as ancilliary files.

\section{Conclusion and Outlook}

In this short paper we discussed an application of the SDE method proposed in~\cite{Papadopoulos:2014lla}. Some partial results of the 
two-loop planar MI for two massless and two general massive external legs with massless propagators were presented. Our full results will be 
published in~\cite{Papadopoulos:2014b}, where the expressions will be added as ancilliary files. All solutions are expressed in terms of 
GPs once the correct $x$-parametrizations of the external momenta are chosen. Furthermore, it was not needed to calculate the boundary terms as they are all captured by the singularities within the SDE method. This seems to be the case for most MI calculated with this method, which makes it an efficient and practical 
tool for the calculation of MI. In the future we intend to apply the method to the calculation of the corresponding non-planar MI counterparts of diboson production. 
No new features are expected to arise for the non-planar graphs and the SDE method should be straightforwardly applicable. In addition, 
the case of massive internal propagators could be the subject of future research.

\providecommand{\href}[2]{#2}\begingroup\raggedright\endgroup


\begin{thebibliography}{10}

\bibitem{Bern:1994zx}
Z.~Bern, L.~J.~Dixon, D.~C.~Dunbar and D.~A.~Kosower, {\em Nucl. Phys.} {\bf B425} (1994) 217-260,
  [\href{http://arxiv.org/abs/hep-ph/9403226}{{\tt hep-ph/9403226}}].
  
\bibitem{Bern:1994cg}
Z.~Bern, L.~J.~Dixon, D.~C.~Dunbar and D.~A.~Kosower, {\em Nucl. Phys.} {\bf B435} (1995) 59-101,
  [\href{http://arxiv.org/abs/hep-ph/9409265}{{\tt hep-ph/9409265}}].
  
\bibitem{Ossola:2006us}
G.~Ossola, C.~G.~Papadopoulos and R.~Pittau, {\em Nucl. Phys.}
  {\bf B763} (2007) 147-169,
  [\href{http://xxx.lanl.gov/abs/hep-ph/0609007}{{\tt hep-ph/0609007}}].

\bibitem{Ossola:2008xq}
G.~Ossola, C.~G.~Papadopoulos and R.~Pittau, {\em JHEP} {\bf 0805} (2008) 004,
  [\href{http://xxx.lanl.gov/abs/0802.1876}{{\tt arXiv:0802.1876}}].
  
\bibitem{vanHameren:2009dr}
A.~van Hameren, C.~Papadopoulos and R.~Pittau, {\em JHEP} {\bf 0909} (2009) 106,
  [\href{http://xxx.lanl.gov/abs/0903.4665}{{\tt arXiv:0903.4665}}].

\bibitem{Berger:2009zg}
C.~Berger, Z.~Bern, L.~J.~Dixon, F.~Febres~Cordero, D.~Forde, et~al., {\em
  Phys. Rev. Lett.} {\bf 102} (2009) 222001,
  [\href{http://xxx.lanl.gov/abs/0902.2760}{{\tt arXiv:0902.2760}}].

\bibitem{Bern:2013gka}
Z.~Bern, L.~Dixon, F.~Febres~Cordero, S.~Hoeche, H.~Ita, et~al., {\em Phys. Rev.}
  {\bf D88} (2013) 014025, [\href{http://xxx.lanl.gov/abs/1304.1253}{{\tt
  arXiv:1304.1253}}].

\bibitem{Bredenstein:2009aj}
A.~Bredenstein, A.~Denner, S.~Dittmaier and S.~Pozzorini, {\em
  Phys. Rev. Lett.} {\bf 103} (2009) 012002,
  [\href{http://xxx.lanl.gov/abs/0905.0110}{{\tt arXiv:0905.0110}}].

\bibitem{Bevilacqua:2009zn}
G.~Bevilacqua, M.~Czakon, C.~Papadopoulos, R.~Pittau and M.~Worek, {\em JHEP} {\bf
  0909} (2009) 109, [\href{http://xxx.lanl.gov/abs/0907.4723}{{\tt
  arXiv:0907.4723}}].

\bibitem{Bevilacqua:2010ve}
G.~Bevilacqua, M.~Czakon, C.~Papadopoulos and M.~Worek, {\em Phys. Rev. Lett.} {\bf 104} (2010)
  162002, [\href{http://xxx.lanl.gov/abs/1002.4009}{{\tt arXiv:1002.4009}}].
  
\bibitem{Gluza:2010ws}
J.~Gluza, K.~Kajda and D.~A.~Kosower, {\em Phys. Rev.} {\bf D83} (2011) 045012,
  [\href{http://xxx.lanl.gov/abs/1009.0472}{{\tt arXiv:1009.0472}}].

\bibitem{Kosower:2011ty}
D.~A.~Kosower and K.~J.~Larsen, {\em Phys. Rev.} {\bf D85} (2012) 045017,
  [\href{http://xxx.lanl.gov/abs/1108.1180}{{\tt arXiv:1108.1180}}].

\bibitem{Mastrolia:2011pr}
P.~Mastrolia and G.~Ossola, {\em JHEP} {\bf 1111} (2011) 014,
  [\href{http://xxx.lanl.gov/abs/1107.6041}{{\tt arXiv:1107.6041}}].

\bibitem{Badger:2012dp}
S.~Badger, H.~Frellesvig and Y.~Zhang, {\em JHEP} {\bf 1204} (2012) 055,
  [\href{http://xxx.lanl.gov/abs/1202.2019}{{\tt arXiv:1202.2019}}].
  
\bibitem{Badger:2013gxa}
S.~Badger, H.~Frellesvig and Y.~Zhang, {\em JHEP} {\bf 1312} (2013) 045,
  [\href{http://xxx.lanl.gov/abs/1310.1051}{{\tt arXiv:1310.1051}}].
  
\bibitem{'tHooft:1978xw}
G.~'t~Hooft and M.~Veltman,  {\em Nucl. Phys.} {\bf B153} (1979) 365-401.
  
\bibitem{Goncharov:1998aa}
A.~B.~Goncharov, {\em Math Res. Letters} {\bf 5} (1998) 497-516,
  [\href{http://arxiv.org/abs/1105.2076}{{\tt arXiv:1105.2076 [math.AG]}}].
  
\bibitem{Remiddi:1999ew}
E.~Remiddi and J.~A.~M.~Vermaseren, {\em Int. J. Mod. Phys.} {\bf A15} (2000) 725-754,
  [\href{http://arxiv.org/abs/hep-ph/9905237}{{\tt hep-ph/9905237}}].

\bibitem{Goncharov:2001aa}
A.~B.~Goncharov, (2001)
  [\href{http://arxiv.org/abs/math/0103059}{{\tt math/0103059v4}}].
  
\bibitem{Kotikov:1990kg}
A.~Kotikov, {\em Phys. Lett.} {\bf B254} (1991) 158-164.

\bibitem{Remiddi:1997ny}
E.~Remiddi, {\em Nuovo Cim.} {\bf A110} (1997) 1435-1452.
  [\href{http://arxiv.org/abs/hep-th/9711188}{{\tt hep-th/9711188}}].

\bibitem{Caffo:1998du}
M.~Caffo, H.~Czyz, S.~Laporta and E.~Remiddi, {\em Nuovo Cim.} {\bf A111} (1998) 365-389,
  [\href{http://arxiv.org/abs/hep-th/9805118}{{\tt hep-th/9805118}}].
  
\bibitem{Gehrmann:1999as}
T.~Gehrmann and E.~Remiddi, {\em Nucl. Phys.} {\bf B580} (2000) 485-518,
  [\href{http://xxx.lanl.gov/abs/hep-ph/9912329}{{\tt hep-ph/9912329}}].
  
\bibitem{Gehrmann:2000zt}
T.~Gehrmann and E.~Remiddi,
  {\em Nucl. Phys.} {\bf B601} (2001) 248
  [\href{http://arxiv.org/abs/hep-ph/0008287}{{\tt hep-ph/0008287}}].
  
\bibitem{Gehrmann:2001ck}
T.~Gehrmann and E.~Remiddi,
  {\em Nucl. Phys.} {\bf B601} (2001) 287
  [\href{http://arxiv.org/abs/hep-ph/0101124}{{\tt hep-ph/0101124}}].
  
\bibitem{Martin:2003qz}
S.~P.~Martin,
  {\em Phys. Rev.} {\bf D68} (2003) 075002
  [\href{http://arxiv.org/abs/hep-ph/0307101}{{\tt hep-ph/0307101}}].
  
\bibitem{Birthwright:2004kk}
T.~G.~Birthwright, E.~W.~N.~Glover and P.~Marquard,
  {\em JHEP} {\bf 0409} (2004) 042
  [\href{http://arxiv.org/abs/hep-ph/0407343}{{\tt hep-ph/0407343}}].
  
\bibitem{Awramik:2004ge}
M.~Awramik, M.~Czakon, A.~Freitas and G.~Weiglein,
  {\em Phys. Rev. Lett.} {\bf 93} (2004) 201805
  [\href{http://arxiv.org/abs/hep-ph/0407317}{{\tt hep-ph/0407317}}].
  
\bibitem{Aglietti:2004ki}
U.~Aglietti, R.~Bonciani, G.~Degrassi and A.~Vicini,
  {\em Phys. Lett.} {\bf B600} (2004) 57
  [\href{http://arxiv.org/abs/hep-ph/0407162}{{\tt hep-ph/0407162}}].
  
\bibitem{Anastasiou:2006hc}
C.~Anastasiou, S.~Beerli, S.~Bucherer, A.~Daleo and Z.~Kunszt,
  {\em JHEP} {\bf 0701} (2007) 082
  [\href{http://arxiv.org/abs/hep-ph/0611236}{{\tt hep-ph/0611236}}].

\bibitem{Bonciani:2008wf}
R.~Bonciani and A.~Ferroglia, {\em JHEP} {\bf 0811} (2008) 065
  [\href{http://arxiv.org/abs/0809.4687}{{\tt arXiv:0809.4687}}].

\bibitem{Henn:2014lfa}
J.~M.~Henn, K.~Melnikov and V.~A.~Smirnov, {\em JHEP} {\bf 05} (2014) 090,
  [\href{http://arxiv.org/abs/1402.7078}{{\tt arXiv:1402.7078}}].
  
\bibitem{Caola:2014lpa}
F.~Caola, J.~M.~Henn, K.~Melnikov and V.~A.~Smirnov, 
  [\href{http://arxiv.org/abs/arXiv:1404.5590}{{\tt arXiv:1404.5590}}].

\bibitem{Papadopoulos:2014lla}
C.~G.~Papadopoulos,
  [\href{http://arxiv.org/abs/arXiv:1401.6057}{{\tt arXiv:1401.6057}}].
  
\bibitem{Papadopoulos:2014b}
C.~G.~Papadopoulos, D.~Tommasini and C.~Wever,
  {{\tt to appear on arXiv}}.
  
\bibitem{Chetyrkin:1981qh}
K.~Chetyrkin and F.~Tkachov, {\em Nucl. Phys.} {\bf B192} (1981)
  159-204.

\bibitem{Tkachov:1981wb}
F.~Tkachov, {\em Phys. Lett.} {\bf B100} (1981)
  65-68.
  
\bibitem{vonManteuffel:2012np}
A.~von Manteuffel and C.~Studerus,
  [\href{http://arxiv.org/abs/arXiv:1201.4330}{{\tt arXiv:1201.4330}}].
  
\bibitem{Smirnov:2008iw}
A.~V.~Smirnov, {\em JHEP} {\bf 0810} (2008) 107.
  [\href{http://arxiv.org/abs/arXiv:0807.3243}{{\tt arXiv:0807.3243}}].

\bibitem{Henn:2013pwa}
J.~M.~Henn, {\em Phys. Rev. Lett.} {\bf 110} (2013) 251601,
  [\href{http://xxx.lanl.gov/abs/1304.1806}{{\tt arXiv:1304.1806}}].
  
\bibitem{Vollinga:2004sn}
J.~Vollinga and S.~Weinzierl, {\em Comput. Phys. Commun.} {\bf 167} (2005) 177.
  [\href{http://arxiv.org/abs/hep-ph/0410259}{{\tt hep-ph/0410259}}].
  
\bibitem{Duhr:2011zq}
C.~Duhr, H.~Gangl and J.~R.~Rhodes, {\em JHEP} {\bf 1210} (2012) 075,
  [\href{http://xxx.lanl.gov/abs/1110.0458}{{\tt arXiv:1110.0458}}].

\bibitem{Duhr:2012fh}
C.~Duhr, {\em JHEP} {\bf 1208} (2012) 043,
  [\href{http://xxx.lanl.gov/abs/1203.0454}{{\tt arXiv:1203.0454}}].

\bibitem{Carter:2010hi}
J.~Carter and G.~Heinrich, {\em Comput. Phys. Commun.} {\bf 182} (2011) 1566-1581,
  [\href{http://xxx.lanl.gov/abs/1011.5493}{{\tt arXiv:1011.5493}}].
  
\bibitem{Borowka:2012yc}
S.~Borowka, J.~Carter and G.~Heinrich, {\em Comput. Phys. Commun.}  {\bf 184} (2013) 396-408,
    [\href{http://arxiv.org/abs/arXiv:1204.4152}{{\tt arXiv:1204.4152}}].


\end{thebibliography}
\end{document}